\documentclass{elsart}

\journal{Journal of Computational Physics}

\usepackage{amssymb}
\usepackage{graphicx}

\begin{document}

\begin{frontmatter}



\title{Parallel implementation of the recursive Green's function method}


\author[TC,IWR]{P.S. Drouvelis\corauthref{cor}}, 
\corauth[cor]{Corresponding author.}
\ead{panos@tc.pci.uni-heidelberg.de}
\author[TC,PI]{P. Schmelcher},
\ead{peter@tc.pci.uni-heidelberg.de}
\author[IWR]{P. Bastian}
\ead{Peter.Bastian@iwr.uni-heidelberg.de}

\address[TC]{Theoretische Chemie, Universit\"at Heidelberg,
Im Neuenheimer Feld 229, D-69120 Heidelberg, Germany}

\address[IWR]{Interdisziplin\"ares Zentrum f\"ur Wissenschaftliches Rechnen,
Im Neuenheimer Feld 368, D-69120 Heidelberg, Germany}

\address[PI]{Physikalisches Institut, Philosophenweg 12, Universit\"at Heidelberg,
D-69120 Heidelberg, Germany}

\begin{abstract}

A parallel algorithm for the implementation of the recursive Green's function technique, which is extensively applied
in the coherent scattering formalism, is developed. The algorithm performs a domain decomposition of the
scattering region among the processors participating in the computation and calculates
the Schur's complement block in the form of distributed blocks among the processors.
If the method is applied recursively, thereby eliminating the processors cyclically, it is possible
to arrive at a Schur's complement block of small size and compute the desired block
of the Green's function matrix directly. The numerical complexity due to the longitudinal dimension
of the scatterer scales linearly with the number of processors, though, the computational cost
due to the processors' cyclic reduction,
establishes a bottleneck to achieve efficiency $100\%$. The proposed algorithm
is accompanied by a performance analysis for two numerical benchmarks, in which the dominant sources of
computational load and parallel overhead as well as their competitive role in the efficiency
of the algorithm will be demonstrated.

\end{abstract}

\end{frontmatter}

% main text
\section{Introduction}

In 1965, Gordon Moore predicted that the number of transistors packed on a chip would continue
to double every year, a prediction known as the Moore's law \cite{Moore65}.
During the past few decades, the rapid progress of novel experimental techniques, has
resulted in scaling down the size of the integrated circuits on a chip
according to Moore's law. Nowadays, chips with hundreds of millions of transistors have been industrially
realised and are exploited in a wide range of commercial applications. However, the continuous scaling
down in size of the transistors is about to reduce their dimensions thereby entering the mesoscopic regime,
in which the electronic wave length
becomes important, i.e., comparable with the size of the device, and quantum effects dominate and define
the laws of information processing \cite{Datta95}. The natural route towards future electronics is therefore to
understand these effects and comprehend them in the design of the nanoscaled devices.
The necessary condition for the description of these phenomena in realistic mesoscopic transistors,
regarding computational resources, is the ability to treat systems with million degrees of freedom.

The theoretical framework for the description of mesoscopic electronic transport has been
established within the Landauer formalism \cite{Landauer89}, in which the conductance of a
mesoscopic sample is in direct relation to the probability that an electron will transmit through it.
To this end, several numerical techniques have been developed and applied to describe various
physical setups. The most efficient method
to attack coherent ballistic transport has proven to be the recursive Green's function (RGF) approach. 
The general framework for this approach can be found in Refs. \cite{Ferry97,Sanvito99,Fagas00} and
depending  on the emphasis of the individual scattering problem, alternative numerical techniques
can be applied. Therefore, techniques such as the boundary element method \cite{Frohne89}, with an emphasis
on the arbitrary geometry of the scattering region, or the modular Green's function method \cite{Rotter00},
in which the scattering region is initially decomposed in modules which are finally joined
via the Dyson equation, have been developed to take into account the particular geometrical features of the
scattering problem. Recently, a RGF technique has been
applied to describe scanning probe experiments \cite{Metalidis04}. This technique describes
tunneling, through the STM tip, which comprises the whole scattering area but scales equally well
with the standard RGF method. As an alternative solution to
improve the efficiency and consequently the capability to treat larger systems,
approximations in the Schr\"odinger eigenvalue problem, as in the contact block reduction method \cite{Mamaluy03},
have been employed to treat multi-terminal three-dimensional problems with relatively good accuracy.

The aim of this paper is to present a parallel algorithm for the computation of the electronic
transmission probability, within the framework of the RGF method.
The parallelization will allow us to treat large systems with millions of degrees of freedom
and will be particularly efficient to handle highly complex modular scattering structures.
The paper is organised as follows. In section \ref{sec2} we discuss
the basic guidelines of the coherent scattering formalism and the desired computational goal to
be achieved. In section \ref{sec3} we construct the parallel algorithm and calculate its numerical
complexity. Section \ref{sec4} contains an analysis of the performance
and scalability of the applied parallel algorithm for
certain numerical benchmarks. Finally, section \ref{sec5} draws our main conclusions.

\section{Basic guidelines of the coherent scattering formalism}
\label{sec2}

\begin{figure*}[h]
\begin{center}
\includegraphics[width=7cm]{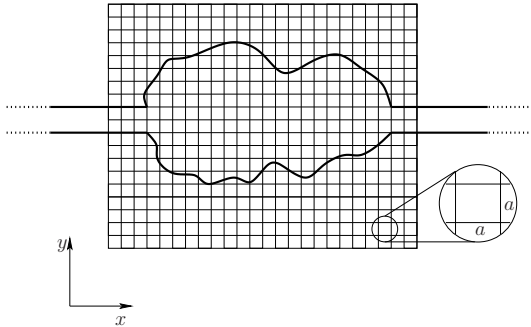}
\end{center}
\caption{\label{fig1} Two-dimensional scattering region attached to two
reservoirs discretized on a uniform lattice with constant $a$.}
\end{figure*}

Coherent scattering formalism implies that the conductance of a mesoscopic sample
attached to two reservoirs (Fig. \ref{fig1}) is proportional to the quantum-mechanical probability $T(E)$ that an
incoming electron at a Fermi energy $E$ in the reservoirs will
transmit through it. To evaluate the transmission probability $T(E)$ one has to
solve the Schr\"odinger equation:

\begin{equation}
(E-H(\mathbf{r})+i\eta) G^R({\bf r};{\bf r'}) = \delta({\bf r} - \bf{r'})
\label{eq1}
\end{equation}

where $H({\bf r})$ is the Hamiltonian and $G^R({\bf r};{\bf r'})$ is
the retarded Green's function operators of the open system (scatterer + reservoirs).
In the following we restrict ourselves to two-dimensional ($2D$) transport.

To proceed with the calculation of $T(E)$ we discretize our space on a uniform lattice with constant $a$.
In order to represent the Hamiltonian operator $H({\bf r})$ we use the tight-binding model assuming
only nearest neighbor interactions \cite{Ferry97}. In this case the Hamiltonian can be written:

\begin{equation}
{\mathbf{H}}({\mathbf{r}}) = \sum\limits_{\mathbf{r}} | \mathbf{r} \rangle \epsilon_{\mathbf{r}} \langle \mathbf{r} |
+ \sum\limits_{\mathbf{r},\mathbf{\Delta r}}
|\mathbf{r} \rangle V_{\mathbf{r},\mathbf{\Delta r}} \langle \mathbf{r}+\mathbf{\Delta r}|
\label{eq2}
\end{equation}

where $\epsilon_{\mathbf{r}}$ is the on-site energy at the position ${\mathbf{r}} = (x,y)$ with $x = na$ and
$y = ma$, $n,m \in \Nset$, $\mathbf{\Delta r}$ represents the vectors
from ${\mathbf{r}}$ to their nearest neighbor sites and $V_{\mathbf{r},\mathbf{\Delta r}}$ is the nearest
neighbor hopping energy. The dispersion relation for the $2D$ discretized lattice reads for a constant
on-site energy:

\begin{equation}
E_{2D}({\mathbf{k}}) = 4V_0 - 2V_0cos(k_x a) - 2V_0cos(k_y a)
\label{eq3}
\end{equation}

where $\mathbf{k} = (k_x,k_y)$ is the electron's wavevector and
$V_0=-\hbar^2/(2m^*a^2)$ is the matrix hopping element linking each site to its nearest neighbor. In the
limit $a \rightarrow 0$ we recover the usual parabolic relationship of a free particle
in a continuum space.

The full tight-binding Hamiltonian of the open system (scatterer + reservoirs)
can be then decomposed in the following block form:

\[ \mathbf{H(r)} = \left( \begin{array}{ccc}

{\mathbf{H_L}} & {\mathbf{V_L}} & {\mathcal{O}} \\

{\mathbf{V_L^\dagger}} & {\mathbf{H_S}} & {\mathbf{V_R}} \\

{\mathcal{O}} & {\mathbf{V_R^\dagger}} & {\mathbf{H_R}} \\

\end{array} \right) \]

where the Hamiltonian ${\mathbf{H_S}}$ describes the electronic motion in an arbitrary scattering region
which is coupled to two external reservoirs from the left and right,
via the semi-infinite matrices ${\mathbf{V_L}}$ and ${\mathbf{V_R}}$ respectively.
The Hamiltonian operators ${\mathbf{H_L}}$ and ${\mathbf{H_R}}$ are of infinite size
and describe the electronic flow within the reservoirs.

Following Datta \cite{Datta95} one can accordingly partition the overall retarded Green's function operator of
equation (\ref{eq1}). It is then possible to obtain for the retarded Green's function
operator of the scatterer the following expression,

\begin{equation}
{\mathbf{G}}(E)
= [E {\mathbf{I}} - {\mathbf{H_S}} - {\mathbf{\Sigma_R}}(E) - {\mathbf{\Sigma_L}}(E) ]^{-1} 
\label{eq4}
\end{equation}

which takes into account the effect of the coupling to the reservoirs, via the so called self-energy matrices
$\mathbf{\Sigma_K}(E) = {\mathbf{V_K^\dagger}} {\mathbf{G_K}}(E) {\mathbf{V_K}}$
due to the left ($\mathbf{K}=\mathbf{L}$) and right ($\mathbf{K}=\mathbf{R}$) reservoir.
The function $\mathbf{G_K}$ is the retarded Green's function operator of the reservoir ${\mathbf{K}}$,
i.e., ${\mathbf{G_K}}(E) = [(E+i \eta) {\mathbf{I}} - {\mathbf{H_K}}]^{-1}$.

Due to the tight-binding's model discretization, the space of the scattering region
now consists of $n=1,2\ldots,N$ slices along the $x$-direction
each of which consists of $m=1,2,\ldots,M$ sites along the $y$-direction. The matrix
$\mathbf{A} = E {\mathbf{I}} - {\mathbf{H_S}} - {\mathbf{\Sigma_R}}(E) - {\mathbf{\Sigma_L}}(E)$
we want to invert in order to evaluate $\mathbf{G}(E)$ is a $N \times N$ block tridiagonal matrix \cite{Ferry97}
whose elements are the blocks $\mathbf{A_{ij}}$ each of which is of size $M \times M$:

\begin{center}
\[ {\mathbf{A}} = \left( \begin{array}{ccccccc}
\mathbf{A_{11}} & \mathbf{A_{12}} & {\mathcal{O}} & \cdots & {\mathcal{O}} & {\mathcal{O}} & {\mathcal{O}} \\
\mathbf{A_{21}} & \mathbf{A_{22}} & \mathbf{A_{23}} & \cdots & {\mathcal{O}} & {\mathcal{O}} & {\mathcal{O}} \\
{\mathcal{O}} & \mathbf{A_{32}} & \mathbf{A_{33}} & \cdots & {\mathcal{O}} & {\mathcal{O}} & {\mathcal{O}} \\
\vdots & \vdots & \vdots & \ddots & \vdots & \vdots & \vdots \\
{\mathcal{O}} & {\mathcal{O}} & {\mathcal{O}} & \cdots & \mathbf{A_{N-2,N-2}} & \mathbf{A_{N-2,N-1}} & {\mathcal{O}} \\
{\mathcal{O}} & {\mathcal{O}} & {\mathcal{O}} & \cdots & \mathbf{A_{N-1,N-2}}
& \mathbf{A_{N-1,N-1}} & \mathbf{A_{N-1,N}} \\
{\mathcal{O}} & {\mathcal{O}} & {\mathcal{O}} & \cdots & {\mathcal{O}} & \mathbf{A_{N,N-1}} & \mathbf{A_{N,N}}

\end{array} \right)  \]
\end{center}

The expression for the evaluation of $T(E)$ can be given in a compact form
within the Fisher-Lee relation \cite{Fisher81}:

\begin{equation}
T(E) = Tr[{\mathbf{\Gamma_R}}(E) {\mathbf{G}}(E) {\mathbf{\Gamma_L}}(E) {\mathbf{G^\dagger}}(E) ]
\label{eq5}
\end{equation}

where ${\mathbf{\Gamma_K}}(E) = i [{\mathbf{\Sigma_K}}(E) - {\mathbf{\Sigma_K^\dagger}}(E)] $
is the strength of the coupling of the reservoir ${\bf K}$ to the scatterer. Due to the fact that
the reservoirs are coupled only to the left and right of the scatterer, the blocks that correspond
to the left interface of the scatterer with the lead, i.e. the upper left block
$\mathbf{\sigma_{L}}(E)$ of ${\mathbf{\Sigma_L}}(E)$,
and to the right interface of the scatterer with the lead,
i.e. the down right block $\mathbf{\sigma_{R}}(E)$ of ${\mathbf{\Sigma_R}}(E)$, are the nonzero blocks
of the matrices ${\mathbf{\Sigma_K}}(E)$.
Therefore, the total self-energy due to the right and left reservoir has the following structure:

\begin{center}
\[ {\mathbf{\Sigma_L}}(E) + {\mathbf{\Sigma_R}}(E) = \left( \begin{array}{ccccccc}
{\bf \sigma_{L}}(E) & {\mathcal{O}} & {\mathcal{O}} & \cdots & {\mathcal{O}} & {\mathcal{O}} & {\mathcal{O}} \\
{\mathcal{O}} & {\mathcal{O}} & {\mathcal{O}} & \cdots & {\mathcal{O}} & {\mathcal{O}} & {\mathcal{O}} \\
{\mathcal{O}} & {\mathcal{O}} & {\mathcal{O}} & \cdots & {\mathcal{O}} & {\mathcal{O}} & {\mathcal{O}} \\
\vdots & \vdots & \vdots & \ddots & \vdots & \vdots & \vdots \\
{\mathcal{O}} & {\mathcal{O}} & {\mathcal{O}} & \cdots & {\mathcal{O}} & {\mathcal{O}} & {\mathcal{O}} \\
{\mathcal{O}} & {\mathcal{O}} & {\mathcal{O}} & \cdots & {\mathcal{O}} & {\mathcal{O}} & {\mathcal{O}} \\
{\mathcal{O}} & {\mathcal{O}} & {\mathcal{O}} & \cdots & {\mathcal{O}} & {\mathcal{O}} & {\bf \sigma_{R}}(E)
\end{array} \right)
\]
\end{center}

Due to the above mentioned structure of the self-energy matrices, it becomes clear that only
the upper left block of ${\mathbf{\Gamma_L}}(E)$,
${\mathbf{\gamma_L}}(E) = i({\mathbf{\sigma_L}}(E) - {\mathbf{\sigma_L^\dagger}}(E))$ and the
down right block of ${\mathbf{\Gamma_R}}(E)$,
${\mathbf{\gamma_R}}(E) = i({\mathbf{\sigma_R}}(E) - {\mathbf{\sigma_R^\dagger}}(E))$ are nonzero. 
Hence, the trace of the product of the four matrices
occuring in equation (\ref{eq5}) simplifies to:

\begin{equation}
T(E) = Tr[{\mathbf{\gamma_R}}(E) {\mathbf{G_{1,N}}}(E) {\mathbf{\gamma_L}}(E) {\mathbf{G_{1,N}^\dagger}}(E) ]
\label{eq6}
\end{equation}

Equation (\ref{eq6}) implies that only the upper right block of
the inverse of $\mathbf{A}$, ${\mathbf{A^{-1}_{1,N}}} = {\mathbf{G_{1,N}}}$
is necessary for the evaluation of $T(E)$. The ultimate goal is therefore to compute ${\mathbf{A^{-1}_{1,N}}}$.

\section{The parallel algorithm}
\label{sec3}

\subsection{Prerequisites}

The overall scattering problem, as discussed in section \ref{sec2}, is summarized to a $N \times N$ block
tridiagonal matrix ${\bf A} = E{\bf I} - {\bf H} - {\bf \Sigma_R}(E) - {\bf \Sigma_L}(E)$
of which each block is of size $M \times M$, where the goal
is to compute the upper right block of the inverse of
${\bf A}$, ${\bf A^{-1}_{1,N}}$.

The algorithm that we pursue should possess the following properties:

\begin{enumerate}
\item Storage requirements should be restricted to a small number of blocks
of size $M \times M$.

\item The number of inversions and multiplications of the $M \times M$ blocks ${\bf A_{ij}}$, which scale as $M^3$,
should be proportional to $N$. This corresponds to the numerical
complexity of the sequential RGF technique in the asymptotic limit of large $N$ and $M$:

\[ C_{\mathrm{seq}}(N,M) \approx N M^3 \]

\item Exploit the fact that the matrix ${\bf A}$ is Hermitian, i.e., for the off-diagonal blocks
is claimed that ${\bf A_{ij}^\dagger} = {\bf A_{ji}}$.

\item The algorithm should be parallelizable.
\end{enumerate}

\subsection{Preparations}
\subsubsection{Change of the inverse under permutation}

Let $\mathbf{P_{ij}}$ be an elementary permutation matrix with the following properties:

\begin{enumerate}
\item Set $\mathbf{\tilde{A} = P_{ij} A}$, then $\mathbf{\tilde{A}}$ is identical to $\mathbf{A}$ except that rows
$i$ and $j$ are interchanged.

\item Set $\mathbf{\tilde{A} = A P_{ij}}$, then $\mathbf{\tilde{A}}$ is identical to $\mathbf{A}$ except that columns
$i$ and $j$ are interchanged.

\item $\mathbf{P_{ij}^{T} = P_{ij} = P_{ji}}$.

\item $\mathbf{P_{ij} \cdot P_{ij}^T = I}$, i.e., ${\bf P_{ij}}$ is orthogonal and self-inverse
${\bf P_{ij}} = {\bf P_{ij}^{-1}}$.
\end{enumerate}

We call $\mathbf{P = P_{i_n,j_n} \ldots P_{i_1,j_1}}$ a permutation matrix. Then
$\mathbf{P^{-1}} = {\bf (P_{i_n,j_n} \cdot \ldots \cdot P_{i_1,j_1})^{-1}} = {\bf P_{i_1,j_1}^{-1} \cdot \ldots \cdot
P_{i_n,j_n}^{-1}} = {\bf P_{i_1,j_1} \cdot \ldots \cdot P_{i_n,j_n}} = {\bf P^{T}}$. Now if we apply
row and column permutations to the matrix ${\bf A}$, ${\bf \tilde{A}} = {\bf P A P^{T}}$ then for the inverse
we have that ${\bf \tilde{A}^{-1}} = {\bf (P A P^T)^{-1}} = {\bf P^{-T} A^{-1} P^{-1}}  = {\bf P A^{-1} P^{T}}$.

% Now we have that for any matrix $\mathbf{A}$, the following diagram commutes:

% \[ \begin{array}{ccccc}
%  &              & {\textrm{\scriptsize apply permutation}} & & \\
%  & {\mathbf{A}} & \longrightarrow & \mathbf{\tilde{A} = P A P^{T}} \\
% \begin{array}{l}
% {\textrm{\scriptsize compute}} \\
% {\textrm{\scriptsize inverse}} \\
% {\textrm{\scriptsize of {\bf{A}}}}
% \end{array}     & \Big\downarrow & & \Big\downarrow & \begin{array}{l}
%                                                       {\textrm{\scriptsize compute}} \\
%                                                       {\textrm{\scriptsize inverse}} \\
%                                                       {\textrm{\scriptsize of $\mathbf{\tilde{A}}$}}
%                                                       \end{array} \\
%  & {\mathbf{A^{-1}}} & \longrightarrow & \mathbf{\tilde{A}^{-1} = P A^{-1} P^{T}} \\
%  &                   & {\textrm{\scriptsize apply permutation}} & & 
% \end{array} \]

The above imply the following two alternative paths for the computation of $\mathbf{A^{-1}_{1,N}}$:

(a) Starting from $\mathbf{A}$ we compute the inverse of it. Then by applying the
appropriate row and column permutations, through operation of the permutation matrices,
it is possible to shift the desired upper right block ${\bf A^{-1}_{1,N}}$ in another position of the inverse.
Respectively, the down left block of ${\bf A}$ is also shifted.
This first path can be illustrated graphically as follows:
\[ \begin{array}{ccccc}

\setlength{\unitlength}{1cm}
\begin{picture}(2.5,2.5)
\put(0,-1.5){\framebox(2.5,2.5){$\mathbf{A}$}}
\end{picture}

& \begin{array}{c} {\textrm{\scriptsize compute}} \\
{\textrm{\scriptsize  inverse of {\bf{A}}}} \\ \longrightarrow \end{array}  &

\setlength{\unitlength}{1cm}
\begin{picture}(2.5,2.5)
\put(0,-1.5){\framebox(2.5,2.5){$\mathbf{A^{-1}}$}
\put(0,0){\begin{picture}(0,0)(0.5,-2.13) \linethickness{0.7mm} {\framebox(0.3,0.3){}} \end{picture}}
\put(0,0){\begin{picture}(0,0)(2.5,0.0) {\dashbox{0.05}(0.3,0.3){}} \end{picture}}}
\end{picture}

& \begin{array}{c} {\textrm{\scriptsize apply row/col}} \\
{\textrm{\scriptsize permutations}} \\ \longrightarrow \end{array} &

\setlength{\unitlength}{1cm}
\begin{picture}(2.5,2.5)
\put(0,-1.5){\framebox(2.5,2.5){$\mathbf{PA^{-1}P^T}$}
\put(0,0){\begin{picture}(0,0)(0.5,-0.53) \linethickness{0.7mm} {\framebox(0.3,0.3){}} \end{picture}}
\put(0,0){\begin{picture}(0,0)(0.89,-0.01) {\dashbox{0.05}(0.3,0.3){}} \end{picture}}
\put(-0.89,0.0){\line(0,1){2.5}}
\put(-2.5,0.9){\line(1,0){2.5}}}
\end{picture}

\end{array} \]

(b) Alternatively, if we start by applying row and column permutations in the initial matrix $\mathbf{A}$, then we
can shift the upper right block $\mathbf{A_{1,N}}$ into another position. If we compute the inverse of the new matrix
then the desired block $\mathbf{A^{-1}_{1,N}}$ will be located at the same position.
Graphically, this second path implies:
\[ \begin{array}{ccccc}

\setlength{\unitlength}{1cm}
\begin{picture}(2.5,2.5)
\put(0,-1.5){\framebox(2.5,2.5){$\mathbf{A}$}
\put(0,0){\begin{picture}(0,0)(0.5,-2.13) \linethickness{0.7mm} {\framebox(0.3,0.3){}} \end{picture}}
\put(0,0){\begin{picture}(0,0)(2.5,0.0) {\dashbox{0.05}(0.3,0.3){}} \end{picture}}}
\end{picture}

& \begin{array}{c} {\textrm{\scriptsize apply row/col}} \\
{\textrm{\scriptsize permutations}} \\ \longrightarrow \end{array} &

\setlength{\unitlength}{1cm}
\begin{picture}(2.5,2.5)
\put(0,-1.5){\framebox(2.5,2.5){$\mathbf{\tilde{A}}$}
\put(0,0){\begin{picture}(0,0)(0.5,-0.53) \linethickness{0.7mm} {\framebox(0.3,0.3){}} \end{picture}}
\put(0,0){\begin{picture}(0,0)(0.9,0.0) {\dashbox{0.05}(0.3,0.3){}} \end{picture}}
\put(-0.9,0.0){\line(0,1){2.5}}
\put(-2.5,0.9){\line(1,0){2.5}}}
\end{picture}

& \begin{array}{c} {\textrm{\scriptsize compute}} \\
{\textrm{\scriptsize  inverse of $\mathbf{\tilde{A}}$}} \\ \longrightarrow \end{array} &

\setlength{\unitlength}{1cm}
\begin{picture}(2.5,2.5)
\put(0,-1.5){\framebox(2.5,2.5){$\mathbf{\tilde{A}^{-1}}$}
\put(0,0){\begin{picture}(0,0)(0.5,-0.53) \linethickness{0.7mm} {\framebox(0.3,0.3){}} \end{picture}}
\put(0,0){\begin{picture}(0,0)(0.9,0.0) {\dashbox{0.05}(0.3,0.3){}} \end{picture}}
\put(-0.9,0.0){\line(0,1){2.5}}
\put(-2.5,0.9){\line(1,0){2.5}}}
\end{picture}

\end{array} \]

Therefore, the diagram implies that {\it computation of the
desired block of the inverse matrix $\mathbf{A^{-1}_{1,N}}$
by following path (a) is equivalent to the computation of $\mathbf{A^{-1}_{1,N}}$ by following path (b) }.

\subsubsection{Expression of the inverse via the Schur complement}

Let any matrix $\mathbf{A}$ with a general $2 \times 2$ block structure:

\[ \mathbf{A} = \left( \begin{array}{cc}

\mathbf{A_{11}} & \mathbf{A_{12}} \\

\mathbf{A_{21}} & \mathbf{A_{22}}

\end{array} \right) \]

Then the inverse of $\mathbf{A}$ in block form is:

\begin{center}

\[ \mathbf{A^{-1}} = \left( \begin{array}{cc}

\mathbf{A_{11} + A_{11}^{-1}A_{12}S^{-1}A_{21}A_{11}^{-1}} \quad & \quad \mathbf{-A_{11}^{-1}A_{12}S^{-1}} \\

\mathbf{-S^{-1}A_{21}A_{11}^{-1}} & \mathbf{S^{-1}}

\end{array} \right) \]

\end{center}

where $\mathbf{S = A_{22} - A_{21} A_{11}^{-1} A_{12}}$ is the so called Schur's complement block.

Hence, together with the permutation Lemma (section 3.2.1) we arrive at the following statement:

{\it If the block ${\bf A_{1,N}}$ is transfered to the block ${\bf A_{22}}$ via permutation transformation
then the desired block ${\bf A_{1,N}^{-1}}$ of the inverse can be obtained from the inverse ${\bf S^{-1}}$ of ${\bf S}$}.

\subsection{Parallel recursive algorithm}

\begin{figure*}[h]
\begin{center}
\includegraphics[width=9cm]{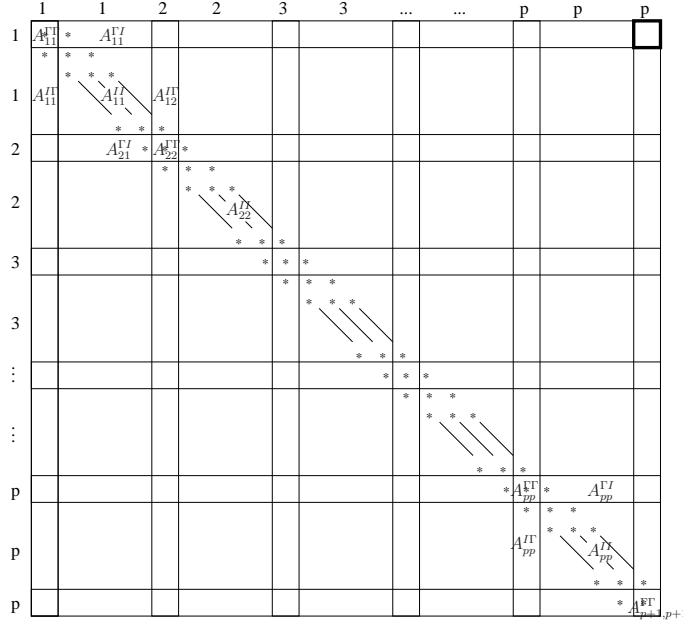}
\end{center}
\caption{\label{fig2}Original block tridiagonal matrix with new secondary level block structure
due to processor subdivision.}
\end{figure*}

To construct the parallel recursive algorithm for the computation of $\mathbf{A_{1,N}^{-1}}$ we proceed as follows.
By starting from the matrix $\mathbf{A}$ in its original block tridiagonal form, we induce an additional
block structure thereby distributing the domains of the scattering region to $p$ processors as shown in
Figure \ref{fig2}. This secondary level block structure, due to the scatterer's domain decomposition, consists
of $p$ blocks, which in turn contain $n_1,n_2,\ldots,n_p$ blocks respectively and $p+1$ elementary blocks which
correspond to the interface slices of the decomposed domains. Additionally, we encounter blocks that couple
the interface slices with the $p$ blocks corresponding to the scatterer's domains. The position of the upper right block
$\mathbf{A^{-1}_{1,N}}$ that is required to be computed is indicated in Figure \ref{fig2}.

\begin{figure*}[h]
\begin{center}
\includegraphics[width=9cm]{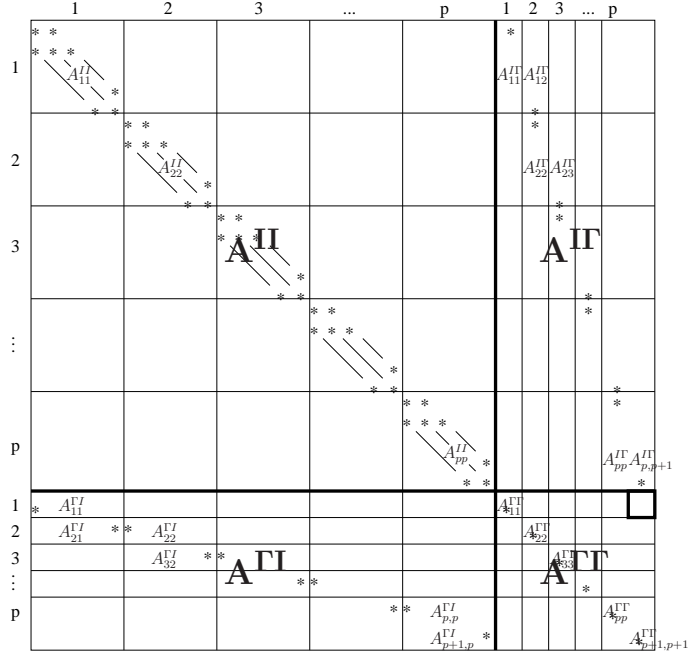}
\end{center}
\caption{\label{fig3}Reordered matrix ${\bf A}$ after row and column permutations.}
\end{figure*}

In the next step we reorder rows and columns, formally through permutation matrices, and we arrive at the
reordered matrix with the structure of Figure \ref{fig3}. The reordered matrix has the $2 \times 2$ block
structure,

\begin{center}

\[ \mathbf{\tilde{A}} = \left( \begin{array}{cc}

\mathbf{A^{II}} & \mathbf{A^{I \Gamma}} \\

\mathbf{A^{\Gamma I}} & \mathbf{A^{\Gamma \Gamma}}

\end{array} \right) \]

\end{center}

and moreover, the desired block to be computed is transfered to the upper right corner of $\mathbf{A^{\Gamma \Gamma}}$.
Therefore, in order to compute $\mathbf{A^{-1}_{1,N}}$, it suffices to compute
$\mathbf{S = A^{\Gamma \Gamma} - A^{\Gamma I}(A^{II})^{-1} A^{I \Gamma}}$ and
extract the upper right block of $\mathbf{S^{-1}}$. The computation of $\mathbf{S}$ results
again in a block tridiagonal matrix and the algorithm can be applied recursively, i.e., by
knowing $\mathbf{S}$ and applying cyclic reduction
among the processors which participate in $\mathbf{S}$, we can arrive recursively to a matrix that
is small enough to compute $\mathbf{A^{-1}_{1,N}}$ directly.

Analytically, the stages to which the parallel RGF algorithm is divided
as well as the corresponding numerical complexities are the following:

\begin{enumerate}

\item First Stage: {\it Scatterer's domain decomposition and computation of $\mathbf{S}$}

\noindent The scatterer is decomposed in domains with $n_1,n_2,\ldots,n_k,\ldots,n_p$ blocks.
Each domain corresponds to one of the altogether $p$ processors participating in the computation
and additionally, there are $p+1$ elementary blocks of each interface slice between the domains
(Fig. \ref{fig2}). At this point we have to note that the last processor stores
two interface blocks, i.e.,  ${\bf A^{\Gamma \Gamma}_{pp}}$ and ${\bf A^{\Gamma \Gamma}_{p+1,p+1}}$.
Then we reorder rows and columns such that the matrix ${\bf A}$ has the block structure of Fig. \ref{fig3}.
Subsequently, the algorithm performs a block Gaussian elimination adapted to the special sparse block
structure of Fig. \ref{fig3}, i.e., it proceeds by eliminating ${\bf A^{\Gamma I}}$ using ${\bf A^{I I}}$.
Analytically, the steps of the block Gaussian elimination applied hereby:

\newcommand{\keyw}[1]{{\bf #1}}
\begin{tabbing}
\quad \=\quad \=\quad \kill
$\forall$ processor $k$ \\
\>\keyw{for} $i = 1 \ldots n_k$ \\
\> \> ${\bf B} = {\bf (A^{II}_{kk})_{ii}^{-1}}$\\
\> \> ${\bf (A^{II}_{kk})_{i+1,i+1}} = {\bf (A^{II}_{kk})_{ii}} - {\bf (A^{II}_{kk})_{i,i+1}^{\dagger}}
{\bf B} {\bf (A^{II}_{kk})_{i,i+1}}$ \\
\> \> ${\bf (A^{I\Gamma}_{kk})_{i+1}} = - {\bf (A^{II}_{kk})_{i,i+1}^{\dagger}}
{\bf B} {\bf (A^{I\Gamma}_{kk})_{i}}$ \\
\> \> ${\bf (A^{\Gamma \Gamma}_{kk})_{i+1}} = {\bf (A^{\Gamma \Gamma}_{kk})_{i}} - {\bf (A^{I\Gamma}_{kk})_{i}^{\dagger}}
{\bf B} {\bf (A^{I\Gamma}_{kk})_{i}}$ \\
\> \keyw{if} $i == n_k + 1$ \\
\> \> ${\bf B} = {\bf (A^{II}_{kk})_{n_k,n_k}^{-1}}$\\
\> \> ${\bf A^{\Gamma \Gamma}_{kk}} = {\bf (A^{\Gamma \Gamma}_{kk})_{n_k}} - {\bf (A^{I\Gamma}_{kk})_{n_k}^{\dagger}}
{\bf B} {\bf (A^{I\Gamma}_{kk})_{n_k}}$\\
\> \> ${\bf A^{\Gamma \Gamma}_{k,k+1}} = - {\bf (A^{I\Gamma}_{kk})_{n_k}^{\dagger}} {\bf B} {\bf A^{I\Gamma}_{k,k+1}}$
\end{tabbing}

\noindent The numerical cost for each processor
scales with $n_p$ inversions of $M \times M$ blocks and requires $6 \cdot n_p$ multiplications of matrices
(see the algorithm above).
With respect to the storage only a few auxiliary blocks of size $M \times M$, independent of $n_k$, are required.
Hence, each processor at the end of the first stage of the computation has stored the diagonal
${\bf A^{\Gamma \Gamma}_{k,k}}$ and off-diagonal ${\bf A^{\Gamma \Gamma}_{k,k+1}}$
block of the Schur complement. At this point we note that the notation used in the subscript of
the blocks of ${\bf S}$ is identical to the one of the blocks of ${\bf A^{\Gamma \Gamma}}$ for convenience.
The last processor computes, in addition to the two previously mentioned
blocks, the last block ${\bf A^{\Gamma \Gamma}_{p+1,p+1}}$.
The numerical complexity for each processor scales, in the limit of large $N$ and $M$, with:

\[
C_1 \approx 7 n_k M^3 \approx 7 \frac{N}{p} M^3
\]

\noindent After the completion of the first stage, the Schur's complement block ${\bf S}$ has been
computed with the form of distributed blocks ${\bf A^{\Gamma \Gamma}_{k,k}}$ and ${\bf A^{\Gamma \Gamma}_{k,k+1}}$
among the processors participating in the computation.
Moreover, ${\bf S}$ has a block tridiagonal structure and is Hermitian:
\hfill\\
\begin{center}
\[ {\bf S} =
\left( \begin{array}{ccccccc}
\mathbf{A^{\Gamma \Gamma}_{11}} & \mathbf{A^{\Gamma \Gamma}_{12}} & {\mathcal{O}} & \cdots & {\mathcal{O}} & {\mathcal{O}} & {\mathcal{O}} \\
\mathbf{A^{\Gamma \Gamma^\dagger}_{12}} & \mathbf{A^{\Gamma \Gamma}_{22}} & \mathbf{A^{\Gamma \Gamma}_{23}} & \cdots & {\mathcal{O}} & {\mathcal{O}} & {\mathcal{O}} \\
{\mathcal{O}} & \mathbf{A^{\Gamma \Gamma ^\dagger}_{23}} & \mathbf{A^{\Gamma \Gamma}_{33}} & \cdots & {\mathcal{O}} & {\mathcal{O}} & {\mathcal{O}} \\
\vdots & \vdots & \vdots & \ddots & \vdots & \vdots & \vdots \\
{\mathcal{O}} & {\mathcal{O}} & {\mathcal{O}} & \cdots & \mathbf{A^{\Gamma \Gamma}_{p-1,p-1}} & \mathbf{A^{\Gamma \Gamma}_{p-1,p}} & {\mathcal{O}} \\
{\mathcal{O}} & {\mathcal{O}} & {\mathcal{O}} & \cdots & \mathbf{A^{\Gamma \Gamma ^\dagger}_{p-1,p}}
& \mathbf{A^{\Gamma \Gamma}_{p,p}} & \mathbf{A^{\Gamma \Gamma}_{p,p+1}} \\
{\mathcal{O}} & {\mathcal{O}} & {\mathcal{O}} & \cdots & {\mathcal{O}} & \mathbf{A^{\Gamma \Gamma ^\dagger}_{p,p+1}} & \mathbf{A^{\Gamma \Gamma}_{p+1,p+1}}
\end{array} \right) \]
\end{center}
\hfill\\
\item Second Stage: {\it Cyclic reduction of the processors participating in the Schur's complement block}

\noindent To proceed further, we exploit the block tridiagonal structure of ${\bf S}$. To this end
we apply a recursive technique called cyclic reduction \cite{Velde94}. The implementation of this technique
requires successive reordering of the processors in such a way that in each step the Schur's complement
block is half the size as before. The first step of the cyclic reduction algorithm is shown in 
Fig. \ref{fig4}.

\begin{figure*}[h]
\begin{center}
\includegraphics[width=10cm]{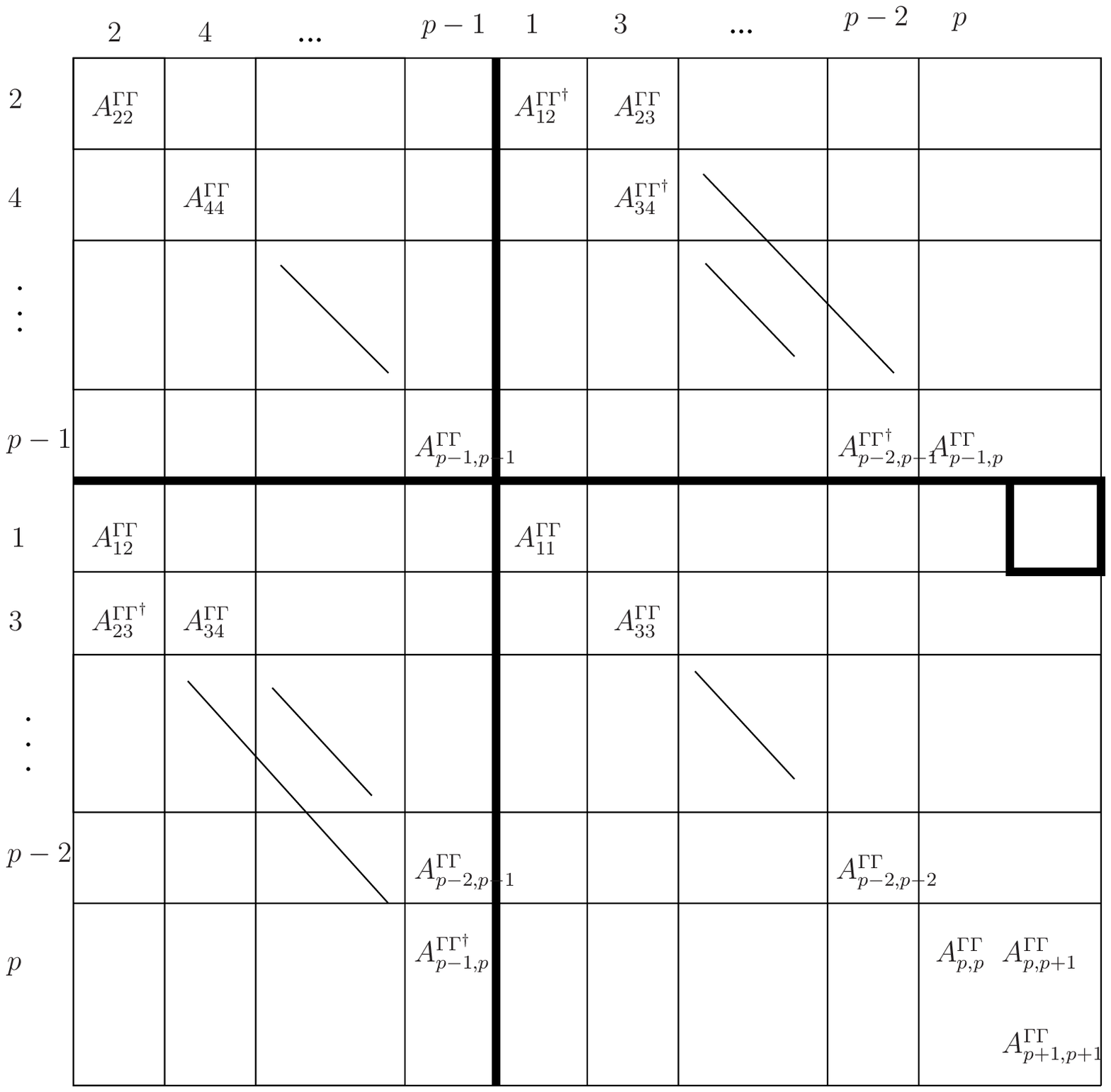}
\end{center}
\caption{\label{fig4} Reordering according to the cyclic reduction algorithm for a Schur's complement block of size
$(p+1) \times (p+1)$. The size of ${\bf S}$ after the applied block Gaussian elimination
is reduced to half of the preceding size.}
\end{figure*}

We observe that the reordered block structure possesses again the $2 \times 2$ structure of the matrix ${\bf {\tilde A}}$.
Therefore by eliminating the off-diagonal block using the upper-diagonal, i.e., the procedure of the first stage,
we arrive at a new Schur's complement block of half the size as the preceding one. By applying this
procedure recursively, after $log_2(p)$ steps we arrive at a $3 \times 3$ block matrix,
of which the upper-right diagonal block of the inverse is the desired ${\bf A_{1,N}^{-1}}$ one.
At this point we should remark that in each recursive step,
the first and the last processor should always participate in the
new resulting Schur's complement block, as shown in Fig. \ref{fig4}. This condition ensures that
the desired block ${\bf A_{1,N}^{-1}}$ is always located in the upper right corner of ${\bf S}$.
In this second stage of parallelization, each recursive step
requires one inversion and four multiplications for the calculation of the diagonal ${\bf A^{\Gamma \Gamma}_{kk}}$
and the fill-in ${\bf A^{\Gamma \Gamma}_{k,k+1}}$ blocks of the resulting Schur's complement block
(see algorithm of the first stage applied to the block structure of Fig. \ref{fig4}).
The numerical complexity of the second stage scales as: 

\[ C_2 \approx 5 log_2(p)  M^3 \]

\noindent After $log_2(p)$ recursive steps, we are left with a $3 \times 3$ block matrix of which
the first row, i.e., blocks ${\bf A^{\Gamma \Gamma}_{1,1}}$ and ${\bf A^{\Gamma \Gamma}_{1,2}}$,
are stored in the first processor and the rest two rows, i.e., blocks ${\bf A^{\Gamma \Gamma}_{p,p}}$,
${\bf A^{\Gamma \Gamma}_{p,p+1}}$ (second row) and ${\bf A^{\Gamma \Gamma}_{p+1,p+1}}$ (third row), are stored in the
last processor. The upper right block of the inverse of this $3 \times 3$ block matrix is the desired
${\bf A_{1,N}^{-1}}$ which can be straightforwardly computed.

\item Third Stage: {\it Computation of the transmission coefficient}

\noindent At the last stage, there remain a few multiplications $c$ of the blocks that are included inside the Fisher-Lee
relation and are all known for the evaluation of $T(E)$. These operations are performed sequentially from
the first processor. The numerical complexity for this last stage can be evaluated as,

\[ C_3 \approx c M^3 \]

\noindent and since $c$ is a small constant, in the limit of large $N$, $C_3$ can be absorbed in $C_1$.

\end{enumerate}

The numerical complexity of the parallel algorithm scales as:

\begin{equation}
C_{\mathrm{par}}(N,M,p) \approx C_1 + C_2 + C_3 \approx 7 \frac{N}{p} M^3 + 5 log_2 p M^3
\label{eq7}
\end{equation}

and the corresponding sequential ($p=1$) one, as:

\[ C_{\mathrm{seq}}(N,M) \approx 7 N M^3 \]

We should remark that the algorithm developed here holds
equally for scattering regions with complex boundary conditions,
i.e., blocks ${\bf A_{ij}}$ with varying sizes,
and can be generalized to the geometry of $3D$ scatterers in a straightforward manner.

\section{Numerical benchmarks}
\label{sec4}

\subsection{Metrics for the analysis of performance and scalability}
\label{subsec41}

In this section an analysis of the performance and scalability for two specific numerical benchmarks
will be pursued. This is required in order to test the models for the numerical complexity we
derived so far and to demonstrate a measure for the capabilities and optimized
use of the proposed algorithm. To proceed with such an analysis it is necessary to define some characteristic quantities
for our parallel algorithm following Ref. \cite{Kumar94}. Firstly, we define the problem size:

\[ W(N,M) = 7 N M^3 \]

which is the number of numerical operations in the sequential algorithm ($p=1$), i.e., the
RGF approach, and is also equal to the serial run time $T_s$ if a unit of time corresponds to each numerical operation.
The cost of simulating the parallel algorithm on a single processor is:

\[ pT_p(N,M,p) = pC_{\mathrm{par}}(N,M,p) = 7 N M^3 + 5\,p\,log_2(p) M^3 \]

where $T_p$ is the parallel run time corresponding to $C_{\mathrm{par}}(N,M,P)$
if we assume a unit of time for each computational
step. The overhead function $T_0$ of the parallel algorithm is defined as:

\[ T_0 (M,p) = pT_p - W = 5\,p\,log_2(p) M^3 \]

and determines the part of its cost that is collectively spent by all processors compared to
the sequential algorithm. The sources of overhead of a parallel system
can be in general attributed to interprocessor
communication, load imbalance and extra computational time due to a part of the program that is not parallelizable.
In our algorithm the dominant contribution to the overhead results from the amount of operations during the
cyclic elimination of the processors.
The extra computational time required for the evaluation of the Fisher-Lee relation (this is the only not parallelizable
part) can be neglected in the limit of large $N$. As far as load imbalance is concerned, the two numerical benchmarks
to be investigated will show a different significance of this source of overhead. Finally,
we define the efficiency of the parallel algorithm as:

\begin{equation}
F = \frac{W}{pT_p} = \frac{7 N M^3}{p \left (7\frac{N}{p} M^3+5\,log_2(p)M^3 \right )}=\frac{1}{1+\frac{5\,p\,log_2(p)}{7 N}}
\label{eq8}
\end{equation}

From this relation, we conclude that the efficiency is independent of the size of blocks $M$
and depends only on the longitudinal length of the scatterer $N$ and the number of processors $p$
participating in the computation. Moreover, by scaling appropriately $N$ with $p$, it is possible to
maintain the efficiency fixed, a property met in scalable parallel algorithms. From Eq. (\ref{eq8})
we can define the isoefficiency function:

\[ W = K T_0 \]

where $K = F/(1-F)$ is given for a specific $F$. For fixed $K$ we can arrive at the following relation for
$N$ and $K$:

\begin{equation}
N = \frac{5}{7} K p\,log_2(p)
\label{eq9}
\end{equation}

Therefore, our algorithm can be cost-optimal if
we choose $N = \frac{5}{7} K p\,log_2(p)$ and scalable if we increase $N$ with rate
$O(p\,log_2(p))$. On the other hand, for a fixed size problem, i.e., keeping $N$ and $M$
fixed, we observe that the efficiency decreases with increasing $p$
as a consequence of Amdahl's law (see Eq. (\ref{eq8})). Here some final
remarks are in order. In the quantities defined so far, we have assumed lattices of unique size $N \times M$ for
the discretization of the scattering regions (perfectly load balanced problems). In addition,
the time spent for the interprocessor communications is neglected. This is due to the increased granularity
of the block tridiagonal system, resulting in a better efficiency of the parallel algorithm.

\subsection{Sinai billiard in a magnetic field}

\begin{figure*}[h]
\begin{center}
\includegraphics[width=7cm]{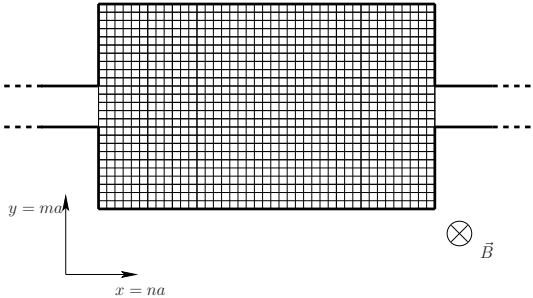}
\end{center}
\caption{\label{fig5} Setup of a Sinai billiard attached to two reservoirs with
$n = 0,1,\ldots,N-1$ slices of $m=0,1,\ldots,M-1$ sites each,
used in the fixed-size efficiency calculations.
The ratio of the two dimensions is $\frac{N}{M} = \frac{8}{5}$.}
\end{figure*}

The first numerical benchmark to test the performance of our algorithm is the Sinai billiard in
a homogeneous magnetic field. Figure \ref{fig5} shows the setup.
Modified Sinai billiard provide a class of systems for testing the
correspondence between quantum and classical transport.
Its simple geometry allows the use of lattices with a unique size
for the discretization, leading therefore to a perfectly balanced problem with respect
to the numerical work loaded to each processor. It represents therefore an excellent ground to test
the models for the complexity we developed in subsection \ref{subsec41}. 

The first setup to test the performance of our algorithm uses a $400 \times 250$ lattice
for the discretization of the Sinai billiard (ten times resolved compared to the one of Figure \ref{fig5}).
The first type of analysis consists of keeping the lattice fixed and studying the efficiency of the problem
with increasing the number of processors. We remind the reader that the total cost of the parallel algorithm
is dominated by the cost for the evaluation of the Schur's complement block and the cost due to the
cyclic reduction of the processors (see Eq. (\ref{eq7})). Table \ref{sinai_a} shows the times measured
for the evaluation of $T(E,B)$ at a fixed energy E and magnetic field $B$. 

\begin{table}[h]
\begin{center}
\caption{Measured time (Time) and efficiency ($F$) as a function of the number $p$ of the processors
for a Sinai billiard in a magnetic field with fixed size $N = 400$ and $M = 250$.}
 \label{sinai_a}
\begin{tabular*}{14cm}{@{\extracolsep{\fill}}ccc|ccc|ccc}
\\
 \hline
 \hline
$p$  &  \hrulefill\ Time (sec) \hrulefill\ & \hrulefill\ $F$ \hrulefill\ &
$p$  &  \hrulefill\ Time (sec) \hrulefill\ & \hrulefill\ $F$ \hrulefill\ & 
$p$  &  \hrulefill\ Time (sec) \hrulefill\ & \hrulefill\ $F$ \hrulefill\ \\

\hline

$1$ & $1723.58$ & $1.0$ & $14$ & $136.82$ & $0.9$ & $48$ & $53.78$ & $0.668$ \\
$2$ & $871.94$ &  $0.989$ & $16$ & $120.09$ & $0.897$ & $56$ & $49.31$ & $0.624$\\
$4$ & $444.75$ & $0.969$ & $20$ & $99.84$ & $0.863$ & $64$ & $45.33$ & $0.594$\\
$6$ & $300.57$ & $0.956$ & $24$ & $86.57$ & $0.83$ & $80$ & $39.68$ & $0.543$ \\
$8$ & $229.18$ & $0.94$ & $28$ & $77.58$ & $0.793$ & $96$ & $38.49$ & $0.466$ \\
$10$ & $185.61$ & $0.928$ & $32$ & $69.51$ & $0.775$ & $112$ & $35.16$ & $0.438$\\
$12$ & $158.46$ & $0.906$ & $40$ & $59.11$ & $0.729$ & $128$ & $34.27$ & $0.393$\\
 \hline
 \hline
 \end{tabular*}
\end{center}
 \end{table}

At this point we note that the system used for the time measurements has been
a Linux cluster of $256$ nodes with Dual AMD Athlon $1.4$ GHz processors of $2$ GB RAM each.
Efficiency is $1.0$ for $p=1$ and gradually decreases
with $p$. This is due to the fact that with increasing $p$, the term
in equation (\ref{eq7}) proportional to $log_2(p)$, dominates
with respect to the other term that decreases with $\frac{N}{p}$,
thereby decreasing the efficiency of the proposed algorithm.

\begin{figure*}[h]
\begin{center}
\includegraphics[width=4cm,angle=-90]{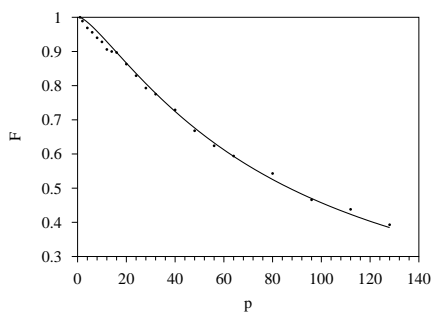}
\end{center}
\caption{\label{fig6} Efficiency $F$ as a function of the number $p$ of processors. The dots correspond to the
measured efficiency and the solid curve to the theoretical model employed.}
\end{figure*}

Figure \ref{fig6} shows the efficiency $F$ as a function of the number $p$ of processors according
to the performed time measurements (dots) compared to the analytical curve of Eq. (\ref{eq8}).
We observe that the agreement between the theoretical model and
the measurements is very good. Therefore, we conclude that the dominant sources of
numerical load have been succesfully identified and weighted. Further sources of overhead, such as the time
required for interprocessors' communication, could be neglected as the work load is dominated by the amount
of numerical operations that scale with $M^3$, i.e., multiplications and inversions of $M \times M$ blocks.

The next step in our analysis is to perform a scaling size experiment. The aim of this test,
is to scale the size of the problem such that the efficiency is kept fixed. As we saw from Eq. (\ref{eq8})
the efficiency is independent of the size of the transversal dimension $M$ and depends only on the size
of the longitudinal dimension $N$ and the number of processors $p$.
Therefore, by scaling appropriately $N$ with $p$
it is possible to arrive at a fixed efficiency $F$ of the algorithm. According to equation (\ref{eq9}) for $p=2$
processors the efficiency can be $0.848$ if we choose $N=8$. If we keep increasing
the number of processors $p$ and the size of the system $N$, keeping $M$ fixed, according to the relation:

\[ N' = N \frac{p'log_2(p')}{plog_2(p)} \]

where $N'$ and $p'$ are the new size of the system and the new number of processors respectively,
then we expect that the efficiency will stabilize around $84.8\%$. Table \ref{sinai_b} shows the
efficiency for the scaled size problem.

\begin{table}[h]
\begin{center}
\caption{Efficiency ($F$) as we increase the longitudinal dimension $N$ of the billiard
with the number of the processors $p$ according to $N = O(p log_2(p))$. We keep $M = 100$ fixed.}
 \label{sinai_b}
\begin{tabular*}{14cm}{@{\extracolsep{\fill}}cccc}
\\
 \hline
 \hline
\hrulefill\ $N;p$ \hrulefill\   &  \hrulefill\ $T_s$ (sec) \hrulefill\
& \hrulefill\ $T_p$ (sec) \hrulefill\ & \hrulefill\ $F$ \hrulefill\ \\

\hline

$8;2$ & $1.1$ & $0.68$ & $0.816$\\
$32;4$ & $4.76$ &  $1.44$ & $0.826$\\
$96;8$ & $14.27$ & $2.17$ & $0.822$\\
$256;16$ & $38.32$ & $2.82$ & $0.849$\\
$640;32$ & $95.25$ & $3.54$ & $0.841$\\
$1536;64$ & $228.79$ & $4.27$ & $0.837$\\
$3584;128$ & $534.06$ & $5.04$ & $0.828$\\
$8192;256$ & $1222.07$ & $5.82$ & $0.82$\\
 \hline
 \hline
 \end{tabular*}
\end{center}
 \end{table}

We observe that the efficiency is stabilized between $0.81$ and $0.85$ thereby confirming our prediction.
The sources of these slight deviations
could be attributed to some enhanced contributions of time spent in interprocessor communications. Therefore
our models provide a reliable source for the estimation of the computational cost. Table \ref{sinai_b}
shows that the larger the size of the system $N$, the larger becomes the efficiency. Therefore,
our parallel algorithm is suitable for large systems, in particular of enhanced longitudinal dimension.
Scattering problems with complex structures could be disentangled into modules with arbitrary
complexity, of which the computation could be done efficiently by one processor. Cyclic reduction
among the processors would join the information of the individual modules. If the computational complexity
of a module is particularly enhanced for one processor, then more processors could be employed.

\subsection{Antidot inside the scatterer}

The second numerical benchmark corresponds to a category of scatterers with enhanced complexity.
It consists of a Sinai billiard with a centered antidot of circular shape.
This setup has been chosen for simulations in Ref. \cite{Fytas05}.
The numerical challenge imposed hereby is the exact reproduction of the antidot's circular shape in the continuum limit.

\begin{figure*}[h]
\begin{center}
\includegraphics[width=14cm]{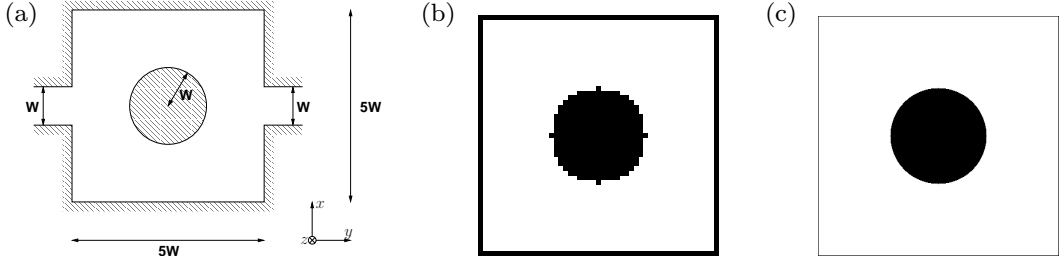}
\end{center}
\caption{\label{fig7} (a) Open scattering geometry of a Sinai billiard
with a centered antidot of circular shape. Subfigure (b) shows the isolated scatterer on a $49 \times 49$
grid of points and width $W = 10a$. Subfigure (c) shows the same setup but four times resolved.
The thickness of the border lines in (b) and (c) provide a measure of the lattice constant.}
\end{figure*}

Figure \ref{fig7} shows the discussed geometry. Subfigure \ref{fig7}-(a) shows the open geometry and dimensions
of the Sinai billiard, while in \ref{fig7}-(b) the isolated Sinai
billiard is discretized on a $49 \times 49$
grid of points. On such a small grid the antidot has, on the scale of Fig. \ref{fig7}-(b), the shape appearance
of a polygon. Subfigure \ref{fig7}-(c) shows the same setup of the Sinai billiard but on a grid which is four times
resolved compared to \ref{fig7}-(b), i.e., a $399 \times 399$ grid.
The latter is going to be our fixed input size for the time measurements
as we increase $p$. At this point we remark that the antidot
has hard wall boundaries, i.e., the sites which form the antidot are excluded from
the computation, thereby leading to blocks ${\bf A_{ij}}$ with varying dimensions. Table \ref{antidot_a}
shows the efficiency measured for the evaluation of $T(E)$ at a fixed energy E as a function of $p$. 

\begin{table}[h]
\begin{center}
\caption{Measured time (Time) and efficiency ($F$) as a function of the number $p$ of the processors
for a Sinai billiard with an antidot placed centrally in it. The lattice $N = 399$ and $M = 399$ is kept fixed.}
 \label{antidot_a}
\begin{tabular*}{14cm}{@{\extracolsep{\fill}}ccc|ccc|ccc}
\\
 \hline
 \hline
$p$  &  \hrulefill\ Time (sec) \hrulefill\ & \hrulefill\ $F$ \hrulefill\ &
$p$  &  \hrulefill\ Time (sec) \hrulefill\ & \hrulefill\ $F$ \hrulefill\ & 
$p$  &  \hrulefill\ Time (sec) \hrulefill\ & \hrulefill\ $F$ \hrulefill\ \\

\hline

$1$ & $13490.83$ & $1.0$ & $14$ & $1201.49$ & $0.802$ & $48$ & $417.8$ & $0.673$ \\
$2$ & $6791.23$ &  $0.993$ & $16$ & $1058.31$ & $0.797$ & $56$ & $379.9$ & $0.634$\\
$4$ & $3917.2$ & $0.861$ & $20$ & $855.45$ & $0.789$ & $64$ & $343.87$ & $0.613$\\
$6$ & $2689.56$ & $0.836$ & $24$ & $734.14$ & $0.766$ & $80$ & $271.07$ & $0.622$\\
$8$ & $1974.65$ & $0.854$ & $28$ & $655.5$ & $0.735$ & $96$ & $267.04$ & $0.526$\\
$10$ & $1649.51$ & $0.818$ & $32$ & $571.54$ & $0.738$ & $112$ & $226.92$ & $0.531$\\
$12$ & $1404.99$ & $0.800$ & $40$ & $462.83$ & $0.729$& $128$ & $224.37$ & $0.47$\\
 \hline
 \hline
 \end{tabular*}
\end{center}
 \end{table}

The efficiency decreases with increasing $p$ as expected. We should note that for these measurements
equidistant domains, with respect to the longitudinal dimension, have been distributed among the processors.
However, due to the antidot's boundaries, it becomes clear
that this kind of distribution leads to an inevitable load imbalance. The
domains that include sections of the antidot are described by blocks of smaller size, resulting thereby
in reduced computational load for the corresponding processors. For $p=2$, we observe an efficiency very close to $100\%$.
This is a result of the symmetry of the geometry of the setup, which results in a load balanced
problem for this specific number of processors. If we further increase $p$ then the efficiency
falls abruptly. This result is attributed to the intensive load imbalance for
few number of processors. To remedy this problem we have to choose a non-uniform
domain decomposition of the scattering region, leading, thereby,
to a more fair work load for all processors. For a larger
number $p$, however, this problem becomes much less intense, since the total cost is multiply distributed
in fairly small pieces of numerical load and the inequality among the processors, with respect
to the load they share, significantly reduces. Therefore, for rather large $p$, load imbalance is
not a significant source of parallel overhead, however, deviations compared to a
load balanced setup are still evident (see below).

To analytically calculate the efficiency of the parallel algorithm for the setup in discussion, it is
necessary to take into account the circular shape of the antidot.
For this purpose, we divide the scatterer in two sections. One section of which the numerical cost scales with
$N_1 \times M^3$ arithmetic operations, where $N_1$ the number of slices outside the antidot,
and a second one of which its computational load scales with $\sum\limits_{i=1}^{N_2} M_i^3$
where $M_i$ is the varying size of the blocks of each of the $N_2$ slices that compose the antidot.
Therefore, the size of the scattering problem is:

\[ W(N,M) = 7 N_1 M^3 + 7 \sum_{i=1}^{N_2} M_i^3 \]

Moreover, we assume that at the first stage of parallelization, the work $W$ is distributed
uniformly among the processors and that at the second stage
the processors that participate in the cyclic reduction are weighted appropriately,
with respect to the load that corresponds to them. This is translated to the fact that $\frac{2}{5}p$ processors
possess a work load that scales with $\sum\limits_{i=1}^{N_2} M_i^3$ and $\frac{3}{5}p$ processors
possess a work load that scales with $M^3$. Therefore, the cost for the parallel algorithm will be:

\[ pT_p=7 N_1 M^3+7 \sum\limits_{i=1}^{N_2} M_i^3 +3plog_2(\frac{3p}{5}) M^3+2plog_2(\frac{2p}{5})\sum_{i=1}^{N_2} M_i^3\]

The efficiency, which is no longer independent of the size of the transversal dimension $M$, will be:

\begin{equation}
F = \frac{W}{pT_p} = \frac{N_1 M^3 + \sum\limits_{i=1}^{N_2} M_i^3}{N_1 M^3+\sum\limits_{i=1}^{N_2} M_i^3 +\frac{3}{7}p\,log_2(\frac{3p}{5}) M^3+\frac{2}{7}p\,log_2(\frac{2p}{5})\sum\limits_{i=1}^{N_2} M_i^3}
\label{eq10}
\end{equation}

\begin{figure}[h]
\begin{center}
\includegraphics[width=4cm,angle=-90]{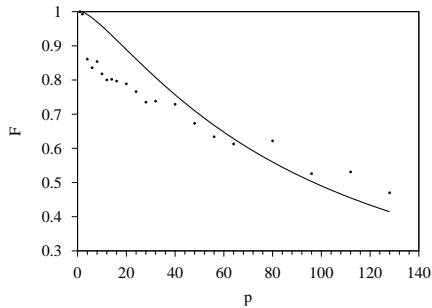}
\end{center}
\caption{\label{fig8} Efficiency $F$ as a funtion of the number $p$ of processors. The dots correspond to the
measured efficiency and the solid curve to the theoretical model derived to take into account
the special geometry of the setup.}
\end{figure}

Figure \ref{fig8} shows the measured efficiency (dots) as a function of $p$. We observe a rather abrupt decrease
of $F$ for a small number of processors $p>10$ which smoothens for larger $p$. 
The solid curve of Figure \ref{fig8} represents the analytical model of Eq. (\ref{eq10}), calculated
for the $399 \times 399$ grid of subfigure (\ref{fig7})-c.

The agreement with the measurements is quite well, however, deviations for $p>2$ are evident. For $p=2$
the prediction agrees due to the symmetric load share to the two processors for this problem.
For $p>2$, deviations are apparent due to the assumptions within the derivation of our model.
Namely, neither does $W$ distribute itself evenly among the processors (load imbalance) nor
is the computational load due to the cyclic reduction weighted exactly among the
processors, as we assumed. To remove the first
assumption one should proceed to an uneven domain decomposition with respect to the processors,
which would vary depending on $p$.
We conclude thereby, that in a scattering problem of complex geometry,
the strategy to be followed in order to optimize the efficiency of the algorithm,
regarding the load that the processors share, should take into account the particular geometric features
of the scatterer.

\section{Conclusions}
\label{sec5}

A parallel algorithm for the implementation of the RGF method has been developed.
The structure of the algorithm is mainly based on an initial domain decomposition of the scattering region
due to processors' subdivision and recursive computation of the Schur's complement block through
cyclic elimination of the processors. The computational cost due to
the longitudinal dimension of the scattering region scales linearly with $p$.
However, the cost due to the cyclic elimination,
prevents us from achieving an efficiency of $100\%$. To demonstrate the efficiency of the
parallel RGF algorithm, we proceeded with an analysis of the performance, scalability and sources of overhead
for two specific numerical benchmarks. The first numerical benchmark
corresponds to a perfectly load balanced setup, such as a Sinai billiard in a magnetic field, and
the derived model is in very good agreement with the measurements. The second numerical example
contained an additional geometrical challenge, being the exact reproduction of the circular shape of an antidot
with hard wall boundaries in the centre of a Sinai billiard. The computation hereby required
manipulation of blocks with varying sizes leading to a nonuniform numerical load for the processors participating
in the computation. A model adapted to the special geometry of this problem has been
employed, which exhibited its geometric peculiarities and
indicated the additional source of overhead due to load imbalance. The effect of the latter can be reduced
by a selection of non-uniform decomposed domains distributed to the processors, based on the numerical cost.
From our analysis it became apparent that the parallel RGF technique developed here,
is particularly suitable for modular scattering structures of high complexity.
Parallelization in this context gives the freedom to decompose the scatterer into modules,
the computation of each can be efficiently performed by one processor.
The optimized distribution of modules to processors depends on their
individual complexity. In case, their complexity is enhanced, more than
one processors could be employed and the corresponding computational load should be shared according to the
individual features of the module.

\appendix

P.\,S.\,D. and P.\,S. gratefully acknowledge illuminating discussions with G. Fagas. P.\,S.\,D. also
acknowledges financial support from DFG in the
framework of the International Graduiertenkolleg IGK 710 "Complex processes: Modeling, Simulation and Optimization".

% \section{}
% \label{}

\end{document}